\documentclass[a4paper,twoside]{article}
\newcommand{\affil}[1]{$^{\rm #1}$}
\usepackage[authoryear]{natbib}
\bibpunct{(}{)}{;}{a}{}{,}
\usepackage{graphicx}
\date{} 
\title{\large\bf\flushleft Booms and Busts:  the Burstiness of Star 
Formation in Nearby Dwarf Galaxies}
\author{\parbox{\textwidth}{\flushleft
\vspace{-0.5cm}
{\it A.A. Cole\affil{A,B}}\\
\vspace{0.4cm}
{\small \affil{A}\,University of Tasmania, School of Mathematics \& Physics,
  Private Bag 37, Hobart, Tasmania 7001, Australia}\\
{\small \affil{B}\,E-mail: andrew.cole@utas.edu.edu}}}
\begin{document}
\twocolumn[
\begin{changemargin}{.8cm}{.5cm}
\begin{minipage}{.9\textwidth}
\vspace{-1cm}
\maketitle
%
%
\small{\bf Abstract:} In this review I summarise recent advances in our 
understanding of the importance of starburst events to the evolutionary
histories of nearby galaxies.  Ongoing bursts are easily
diagnosed in emission-line surveys, but assessing the timing and intensity
of fossil bursts requires more effort, usually demanding color-magnitude
diagrams or spectroscopy of individual stars.  For ages older than $\sim$1~Gyr,
this type of observation is currently limited to the Local Group and its
immediate surroundings.  However, if the Local Volume is representative
of the Universe as a whole, then studies of the age and metallicity
distributions of star clusters and resolved stellar populations should
give statistical clues as to the frequency and importance of bursts to the
histories of galaxies in general.  Based on starburst
statistics in the literature and synthetic colour-magnitude diagram studies
of Local Group galaxies, I attempt to distinguish between systemic starbursts that
strongly impact galaxy evolution and stochastic bursts that can appear impressive
but are ultimately of little significance on gigayear timescales.
As a specific case, it appears as though IC~10, the only starburst
galaxy in the Local Group, falls into the latter category and is not fundamentally
different from other nearby dwarf irregular galaxies.

\medskip
\medskip
\end{minipage}
\end{changemargin}
]
\small

\section{Introduction}

Under concordance cosmologies, galaxy mergers both 
major and minor are expected to play a central role in the evolution of 
virtually every galaxy.   
Starbursts triggered by
tidal and/or hydrodynamic interactions between galaxies are a major
driver of morphological transformation and, though rare, are prominent
contributors to near- and mid-infrared emission in the Universe.  As
just one example, starbursts
driven by major mergers are thought to be an important driver in the
creation of giant elliptical galaxies \citep{ren06}.

Typical large galaxies generally behave as approximately 
self-regulated systems and exhibit a relatively smoothly 
varying star formation rate that depends on gas density
\citep[e.g.,][]{ken98a}, and is likely to slowly decline with time subsequent
to their initial formation \citep{lar78}. 
Starbursting states in
large galaxies are rare events, likely triggered by mergers or strong
tidal interactions \citep[e.g.,][]{mih96,sam03}, and are 
often easily distinguished from a simple
scaling-up of the global star-formation rate by manifesting as
highly localized nuclear or circumnuclear starbursts
\citep[e.g.,][]{ken98b}.

Conversely, small galaxies
are not expected to show constant or smoothly declining
star formation rates,
because they are
far more susceptible to disruption by internal feedbacks and external
perturbations \citep[e.g.,][]{sti07}. 
This presents an observational challenge, because
only within 10-15~Mpc can individual stars be resolved in order to
study the stellar populations of starbursts in detail, and only within
$\approx$1~Mpc can stars be resolved to ages approaching a Hubble time 
to characterize the underlying stellar populations and search for fossil
bursts.  However, within these distances luminous galaxies are rare, far 
outnumbered by dwarf galaxies.  Thus we have plentiful opportunity to 
study the burstiness of small galaxies in great detail, but the 
significance of observed variations in SFR is complicated by the 
expectation of large random fluctuations intrinsic to small galaxies.

One of the challenges of the studies of dwarf galaxy star-formation 
histories is to distinguish between systemic starbursts that
are qualitatively different from steady
state or quiescent star formation, and stochastic bursts which are merely the
manifestation of normal variation.  Offsetting this difficulty is
the advantage that in resolved stellar populations there is a large
array of tools available to measure the timing, intensity, and
metallicity of star-formation episodes of any age up to a Hubble time.
This means that burstiness studies can be made
of nearby galaxies regardless of their current morphology or gas
content: even galaxies with no future have a history.  The study
of burst histories in early-type galaxies has the potential to 
illuminate the processes of hierarchical assembly of large galaxies
and morphological transformation of galaxies from disk-dominated
to spheroidal \citep[e.g.,][]{may07}.

Star-forming, gas-rich galaxies are of course easier targets for study,
because of the higher light-to-mass ratios of young stars and the 
possibility to identify optically faint systems via HI surveys.
Late-type galaxies in the nearby Universe exhibit a wide range of 
specific star formation rates ranging from nearly inactive to 
extreme starburst conditions \citep{hun86}.  The only galaxies with neutral gas
detections that do not seem to be forming stars are the lowest-luminosity
examples \citep{hun85}, the transition-type dwarfs such as the recently-discovered
Leo~T dwarf \citep{irw07}.  These dwarfs, which include among their number
the Phoenix, Pisces, and Pegasus systems,
are very faint (M$_B > -14$), isolated systems.  They may be showing the
signs of a breathing mode of star formation \citep{sti07},
in effect experiencing an ``anti-starburst'', or they could merely be
forming stars at such a low rate that no star massive enough to ionize
hydrogen has been produced within the past $\approx$10$^7$ years
\citep{lee09}.

In this review paper I will not discuss the characteristics and causes
of ongoing strong starbursts; an excellent summary of the subject can be
found in \citep[e.g.,][]{gal05}.
The purpose of this paper is to review what is known about 
the role of bursts over the lifetimes of nearby galaxies, to discuss the
lines of evidence that could be used to infer the presence of fossil bursts
in resolved stellar populations, and to draw attention to some of the 
recent work on starburst statistics and durations in the Local Volume.  
For a thorough discussion of star formation in all modes and galaxy types,
including the starburst phenomenon in context, see the review by 
\citet{ken98b} and references therein.

Since the majority of the evidence for fossil starbursts must be gleaned
from color-magnitude diagrams (CMDs), the most detailed results are by
necessity confined to the Local Group, galaxies within about $\sim$1~Mpc
of the barycentre of the M31-Milky Way pair.  Within this group of $\approx$50
galaxies, we find examples of late-type galaxies experiencing both booms
and busts in their current specific star-formation rate; of galaxies
that burst during their formation and never again; of galaxies that burst 
multiple times at inteverals of several Gyr; and galaxies for which there
is no evidence of strong variations in SFR at all.  There even appears to
be a galaxy which saved much its gas for 5 billion years after its
first star formation, whereupon it experienced a major SFR event at a lookback
time corresponding to a redshift $z$ $\approx$ 1; there are hints that
this type of star-formation history (SFH) may be commonplace among the most
isolated small galaxies.  Comprehensive reviews
of the SFH of Local Group galaxies are to be found
in \citet{mat98} and \cite{tol09}, among others.

\section{Dwarf Starbursts: Prevalence and Properties}

The archetypal dwarf galaxy starburst is M82 
\citep[][and references therein]{gal99}.  It displays most
of the characteristics associated with an extreme starburst environment,
including the formation of massive super star clusters, strong
infrared emission, remarkably high H$\alpha$ equivalent width,
strong tidal interaction with a massive neighbor, and a galactic
wind driven by the large number of supernovae resulting from the starburst.
This is an incontrovertible example of star formation in a 
dramatically different mode from that experienced by most dwarfs in the 
nearby Universe.  

Quite different, but equally important
examples of starbursts in dwarf galaxies are the blue compact
dwarfs (BCDs)-- the ``extragalactic HII regions'' of \citet{sar70}.  Because
BCDs are a rare galaxy type, they are typically only found at distances 
greater than 10~Mpc, and attention naturally focuses on their highest
surface brightness features.  However, unlike M82, BCDs are not commonly
found to be interacting with massive companions, and may but do not necessarily
show evidence for the formation of massive star clusters or unusually
centrally concentrated star formation \citep[e.g.,][]{alo99}.
This variation of detail
speaks to the probability that star formation in small galaxies is 
subject to stochastic fluctuations, possibly leading to strong differences in 
instantaneous SFR without implying a qualitative difference in the modes,
triggers, and timescales of star formation \citep[e.g.,][]{wei08}.

There has been a tremendous amount of work done in cataloguing the properties
and populations of dwarf starbursts within $\approx$10~Mpc, but completeness 
and homogeneity of data have been high barriers to putting their properties into
context and beginning to assess the importance of various physical processes
at work.  Recently, a number of surveys have come together to scale this barrier,
providing deep and uniform samples of galaxy broadband luminosities and colors,
H$\alpha$ equivalent widths, and neutral gas content
\citep[e.g.,][among others]{sal01,bri04,meu06,ken08,dal09}.
These allow, for the first time, statistically
sound estimates of the fraction of starbursting galaxies, the fraction of 
total star formation that occurs in bursts, and the duration of typical starbursts.

A thorough review of all the recent survey work on dwarf galaxies in the 
Local Volume is far beyond the scope of this paper, so I will focus on 
one set of recent results that bears directly on the burstiness of small
galaxies.
A major result of the 11~Mpc H$\alpha$ UV Galaxy Survey (11HUGS) survey has 
been published by \citet{lee09}, giving star formation rates based on H$\alpha$
luminosities for over 300 galaxies within 11~Mpc, complete down to apparent magnitude
B $\approx$15. This allows a complete census of star fomation rates in small galaxies,
particularly attuned to studies of starburst statistics when a starburst is defined
purely in terms of the ratio of current SFR to lifetime average.  The major findings of 
\citet{lee09} can be summarized as follows:

$\bullet$ dwarf galaxies with H$\alpha$ equivalent widths $>$100\AA\ make up only
6$^{+4}_{-2}$\% of late-type galaxies, and a correspondingly low fraction of
star formation, 23$^{+14}_{-9}$\%, occurs during burst events.

$\bullet$ Non star-forming late-type galaxies are as rare as starbursts.  However,
the galaxies that lack H$\alpha$ emission are all fainter than M$_B < -13.6$.
Therefore sampling effects on the initial mass function mean that small amounts 
of star formation, in keeping with low overall galaxy masses, could be taking
place without necessarily producing any star massive enough to ionize its natal
cloud.

$\bullet$ It is likely that star formation never ceases completely between bursts,
instead falling to a rate $\approx$4 times less than the peak rate, on average.
This conclusion rests on the assumption that every galaxy is equally likely to
become the host of a burst; further work is needed to test this assumption.

The 11HUGS dataset sets a new standard for the statistics of star formation 
properties of nearby dwarf galaxies, and the conclusions of \citet{lee09} 
make a very secure foundation on which to build a comprehensive theory of
star formation in small galaxies. 
The conclusions based on H$\alpha$ data will be
extended in time by the addition of ultraviolet data, owing to the fact that
Balmer continuum-bright B stars live for an order of magnitude longer on
the main-sequence
than do Lyman continuum-bright O stars.

Further extension to the time
baseline can be provided by diffraction-limited imaging that is capable
of resolving individual stars as deeply as signal-to-noise and crowding
permit.  It is the addition of the temporal component for studies of an
individual galaxy that will ultimately provide the complete picture of 
the burstiness of star formation.  Such data have been obtained for Local
Group galaxies and for star-forming galaxies up 
to several Mpc away and ages of up to 10$^9$ yr
\citep[e.g.,][]{can03,mcq09}, with the result that the typical
starburst in dwarf galaxies appears to last for a few times 10$^8$ years.  
During these periods of heightened SFR, the sites of star formation 
move around the galaxy, propagating at speeds of $\sim$10~km~s$^{-1}$
\citep[e.g.,][]{doh98} and producing what would be observable as a 
``flickering'' if the SFR at just one location was to be measured 
\citep{mcq09}.  These independent constraints on the duration and duty cycle
of starbursts provide complementary information to the statistics of current
bursts embodied by the 11HUGS results, and should be strong constraints on
numerical models that attempt to account for fluctuations in SFR in small
galaxies \citep[e.g.,][]{sti07}.

\section{Identifying Starburst Fossils}

If strong starbursts are assumed to account for $\sim$25\% of the lifetime
integrated star formation in dwarf galaxies, then it should be possible to
see the evidence for fossil bursts in the resolved stellar populations of nearby
galaxies.  The most direct way to unearth fossil bursts is via direct probes of
the star formation history of a galaxy through analysis of its color-magnitude
diagram \citep[CMD,][]{tos91,tol96,dol02}.
As a stellar population ages, the absolute magnitude of its
main-sequence turnoff increases, enabling direct tests of stellar mass contained
in a galaxy as a function of age.  
The classical approach to this problem is to overlay theoretical stellar
isochrones of various age and metallicity combinations on the observed CMD and
thereby identify the characteristics of the dominant stellar populations.  
Particularly narrow sequences similar to star cluster CMDs would be indicative of
a burst of star formation, while gaps in the CMD result from quiescent epochs 
in the life of the galaxy.

For most galaxies the process cannot simplified to such an extent, beacuse of the
continuous distribution of stellar ages and metallicities, and the resulting
extremely large number of different isochrone combinations to be tested.  
Quantitative estimates of stellar masses formed are made difficult in composite
populations (i.e., nearly every galaxy in the Universe), because
older populations are masked to some extent by the low-mass members of younger
populations, and metallicity evolution can counteract some of the dimming effects
of age.  The solution to this problem is to compare the observed density of stars
in a CMD to probability distributions formed by the convolution of isochrones
with an initial mass function.  The difference between the synthetic CMDs thus created 
and a given dataset can then be minimized using a nonlinear least-squares approach.
Other properties of the stellar populations, including the dispersion in
interstellar reddening values, proportion of binary stars, and detailed 
elemental abundance ratios (e.g., [$\alpha$/Fe]) may also require modelling in order
to obtain meaningful results.  The coefficients returned by the minimization 
procedure correspond to the star-formation rate as a function of time (and metallicity,
reddening, [$\alpha$/Fe], or any other parameters the investigator thinks their
data can constrain).

Such techniques of CMD fitting have been widely applied throughout the Local
Group over the past 2 decades, and the results have been reviewed comprehensively
by \citet{mat98} and \citet{tol09}, among others.  The field was brought to 
maturity by the diffraction-limited imaging of the Hubble Space Telescope, which
overcomes much of the stellar crowding that plagued ground-based imaging.  It has
been possible with HST to directly measure the SFH of galaxies over their entire
history, with a time resolution of about 10\% of the age, throughout the Local
Group.  Observations of more distant galaxies have been limited to younger lookback
times by the practical limit of HST imaging at magnitude $\approx$29--30 for most
projects.  This means that for some of the most extreme and interesting starbursts,
e.g., M82 (d $\approx$3.9~Mpc, m$-$M$_0$ $\approx$27.9) direct age-dating via 
the main-sequence turnoff is not possible for ages greater than a few times 10$^9$
years.

\subsection{Indirect Probes of Burst Histories}

Current starbursts are visible to great distances, but fossil bursts
begin to fade and rapidly become difficult to accurately characterize.
No better example of the blurring effects of time is available than the
case of the Large and Small Magellanic Clouds.  At a distances of just 48 (LMC) and
55 (SMC) kpc from the Milky Way, and with a center-to-center separation
of 22~kpc, the Magellanic Clouds are obviously and strongly interacting
with the Milky Way and with each other.  However, neither galaxy is 
currently experiencing a starburst, and there has been great debate
in the literature over the degree of burstiness in their past histories.
It is far beyond the scope of this review to discuss all of the 
observational evidence pro and con for a bursting SFH as opposed to
a smooth SFH, but the arguments can be followed in the proceedings of IAU
Symposia devoted to the Magellanic system \citep{chu99,van09}, the monograph by
\citet{wes97}, and the excellent review by \citet{ols96}.   

In short,
the LMC contains several globular clusters like those in the Milky Way,
plus a large number of younger, massive, dense clusters unlike any found in the 
Galaxy.  These young globulars have age distributions peaked at ages of
$\approx$100--200~Myr and $\approx$1--2~Gyr, leading to suggestions that
the LMC must have experienced strong starbursts at those ages. 
The SMC also contains young, massive clusters, and their ages also
appear to be concentrated at specific times \citep{ric00}-- although not
at the same times as the LMC.  CMD studies
of field stars in the LMC based on HST imaging found varying degrees of 
evidence for a bursting SFH, but it became apparent that the LMC never
completely ceased forming stars in the periods between its epochs of 
prolific cluster formation, and these data were used to argue in favor
of a SFH that was more smooth than bursty.  

The most 
complete derivations of the SFH of the Clouds are to be found in the
work of \citet[][SMC]{har04} and \citet[][LMC]{har09}, and 
the results show that the star formation activity in the field was 
indeed peaked during the times of massive cluster formation, and the times
of activity correlate between the two Clouds.  It therefore does appear
that the production of massive star clusters is a signature of major
events in the life of a galaxy, indicative of star formation under
different conditions than normal, ``quiescent'' star formation.  
Following the example of the Magellanic Clouds, the presence of massive
star clusters can be a tracer of fossil starbursts when the evidence
from field star ages is insufficient to unambiguously identify a burst.

The production of massive star clusters is not expected as the result
of the normal, stochastic burstiness one sees in dwarf galaxies.  The 
presence of massive clusters can be diagnostic of bona fide, systemic
starbursts, and if the clusters are dense enough to escape dissolution 
in the tidal field of the galaxy, they can be powerful probes over a long
time baseline.   This is particularly useful because massive clusters
are far more easily observable than even the brightest single supergiant
stars, and can thus trace starbursts to larger distances and more 
crowded environments \citep[e.g.,][]{gal99}.  Super star clusters
are indeed a ubiquitous feature of starburst galaxies, and in some
starbursts the clusters appear likely to survive to high ages and eventually
appear similar to globular clusters \citep[e.g.,][and references therein]{deg01}.

The burstiness of star formation can also have strong implications for
the chemical enrichment of galaxies \citep[e.g.,][1997]{kob96},
and these chemical signatures persist
in the subsequent generations of stars, long after the bursts responsible
have faded away.  Starbursts can rapidly enrich the interstellar medium
of a galaxy, which might otherwise be diluted in metal abundance by the
infall of fresh metal-poor gas during ``quiescent'' star formation.  The
timescales of starbursts can potentially be probed by the measuring the
abundance ratios of elements produced in Type II and Type Ia supernovae,
and in AGB stars \citep[][and references therein]{tol09}.
However, if the starburst is extreme enough to produce a large number
of supernovae in a small galaxy, then the metals produced could be 
lost to the intergalactic medium rather than incorporated into subsequent
stellar generations, reducing the effective metal yield of the population
in a starburst \citep{mac99,lee06}. The
idea of identifying the star-formation environments of stars by identifying
telltale patterns of chemical enrichment has enormous potential for
understanding the formation and evolution of galaxies from the earliest
times to the present day; for a detailed description of the promise and
challenge of this approach, see the review by \citet{fre02}.

\section{Local Group Case Studies}

\subsection{The Carina Dwarf Spheroidal}

The dwarf spheroidal galaxies are low-luminosity (M$_B > -14$), late-type
galaxies that are nearly exclusively observed as satellites of the Milky Way 
(or M31) with distances of 25 $\leq$ r$_{gc}$ $\leq$ 250~kpc \citep{mat98}.
Many of them appear to have only ever experienced one episode of star formation,
at the earliest epochs, leading to much conjecture about the mechanisms for
their gas-loss and their survival (see the discussions in \citet{tol09} and
\citet{mat98} for further information).  Especially among the lower-luminosity
and less distant spheroidals, the star formation was restricted to more or
less old ages \citep[e.g.,][]{mat91,dol02}, although some chemical evolution has
been reported \citep[e.g.,][]{tol04}, indicating some complexity to the 
SFH.  The most luminous dwarf spheroidals, the Fornax and Sagittarius systems,
show both extended SFHs and small globular cluster populations.

Despite the overall trends pointing to old populations and simple SFHs, the 
classic example of a galaxy with a history of repeated burst cycles is a 
dwarf spheroidal, the Carina system.  At a distance of 100~kpc, Carina has
been known to harbor a large intermediate-age population since the work
of \cite{mou83}.
The first imaging to reach the level of the horizontal branch and
helium-burning red clump revealed a surprisingly bimodal stellar distribution 
implying a bursting or gasping SFH \citep{sme94}.  Subsequent deeper imaging
confirmed this and revealed that the dominant stellar population was formed in a burst
about 7--9~Gyr ago, with clearly separated bursts at both older ($>$11~Gyr) and 
younger ($\approx$4~Gyr) ages \citep{hur98}.  Medium-resolution spectroscopy
of individual red giant stars revealed
that the metallicity of the younger stars is on average slightly higher than
that of the older stars, demonstrating chemical evolution with time 
\citep{koc06}.

Carina remains the clearest example of a galaxy in which ancient and 
intermediate-age bursts are so distinctly separated from each other that
the quiescent 
epochs produce gaps in the subgiant region of the CMD.  Qualitatively
similar behavior is seen in several other early- and late-type dwarfs,
e.g., Leo~I \citep{dol02} and IC~1613 \citep{ski03}, but the variations in 
SFR on Gyr timescales appear to be milder than in Carina.  There are several
ways in which to interpret this information. 
It is possible that Carina is just far enough from the Milky Way to have
escaped the early stripping of gas that terminated the star formation of
the closer-in spheroidals like Sextans, Ursa Minor, or Draco \citep{tol09}.
Alternatively, perhaps Carina 
{\it did} consume or eject its entire gas content after an initial burst, but
accreted more gas several gigayears later.  In comparing Carina to more 
distant galaxies, it must be noted that some of the apparent
smoothness in SFH at old ages for galaxies more than a few hundred kiloparsecs
distant may yet be attributable to observational difficulties rather than
genuine constancy of SFR.  Because the fractional age resolution of the
current generation of CMD studies of Local Group dwarfs is at best about
10\%, starbursts of the durations reported by, e.g., \citet{mcq09} would
be unresolved for ages greater than $\sim$10$^{9.5}$ yr, causing galaxy
SFHs to appear to decrease in burstiness with age.

\subsection{A Late-Bursting Galaxy: Leo~A}

Carina and the other early-type, gas-free, dwarf spheroidal systems are nearly
all satellites of the Milky Way, which is expected to have had a strong impact
on their evolution \citep[e.g.,][]{kra04}.  It is therefore of interest to 
study more distant galaxies to similar photometric depth, in order to gauge
the evolution of small galaxies in relative isolation.  Leo~A (DDO~69), is 
one of the most isolated galaxies in the Local Group \citep{mat98}, 800~kpc
from the Milky Way and 1200~kpc from M31.  It was suggested by \citet{tol98}
that Leo~A had formed the vast majority of its stars within the past few Gyr,
making it the most likely candidate yet to be a genuinely young galaxy.  
This candidacy was quashed when \citet{dol02+} discovered a small number of 
RR~Lyrae type variables in Leo~A, proving the presence of ancient (age
$>$10~Gyr) stars.  However, deep HST imaging by \citet{col07} showed conclusively
that Leo~A formed 90\% of its stellar mass more recently than $\approx$7~Gyr
ago, corresponding to a redshift of $z$ $\approx$1.  The SFR appeared to peak
between 1--3~Gyr ago and then declined, with a second episode of star formation
a few hundred Myr ago.  This made Leo~A unique among known galaxies in having
such a remarkably high fraction of stars younger than 10~Gyr.   In the 
SFH derived by \citet{col07}, the average SFR of Leo~A from 2--5~Gyr ago
was 4--5 times the previous long-term average SFR.  If this star formation 
was concentrated into episodes a few hundred Myr in duration, the actual 
instantaneous SFR would have qualified Leo~A as an extreme dwarf starburst.

It is a puzzle how
such a small galaxy (M$_{dyn}$ $\approx$2$\times$10$^8$ M$_{\odot}$) could have 
retained so much of its gas without forming significant amounts of stars for over 
5~Gyr after it first formed.  This may be an example of an HI reservoir at
low metallicity, with correspondingly long cooling timescale, kept in such isolation
that no perturbations or fluctuations triggered star formation for many gigayears.
Based on Leo~A's very small radial velocity of $-$18 km~s$^{-1}$ with respect
to the Milky Way, it seems likely that the galaxy has never been
in close proximity to either M31 or the Milky Way. 
Interestingly, similar hints
of a large fraction of delayed star formation are seen in IC~1613 \citep{ski03}
and DDO~210 \citep{mcc06}, although not to the same extent as in Leo~A.  It may
become apparent with further observational effort that delayed star formation
with a late burst of star formation-- perhaps triggered by the turnaround and
infall of the isolated galaxy into an intragroup medium \citep{mcc07}
is a typical characteristic of the most isolated galaxies.

\subsection{Boom or Bust? The Case of IC~10}

The  gas-rich
dwarf (M$_B = -15.6$) IC~10 has long been considered the only starburst
galaxy in the Local Group and thought of as something of an anomaly
\citep[e.g.][]{hun01}.
IC~10 has an unusually high surface brightness for a galaxy
of its size, and unusually high numbers of Wolf-Rayet stars
\citep{mas98}, owing to its high current SFR of 
$\approx$0.03 M$_{\odot}$~yr$^{-1}$~kpc$^{-2}$ \citep{hun01}.
At a distance of $\approx$800~kpc from the Milky Way (and a mere 250~kpc 
from M31), 
IC~10 is clearly an outlier among the dwarf irregular galaxies of the 
Local Group, but how much of an outlier is still a matter of debate.
It sits deep in the Zone of Avoidance ($\ell = 119^{\circ}$, $b = -3^{\circ}$),
so detailed study has been hindered by large and variable foreground 
reddening compounded with a significant amount of reddening internal to the
dwarf, making it the last star-forming Local Group galaxy within 1~Mpc
to be imaged down
to the depth of the horizontal branch \citep{san09}.  Attention has naturally
been focused on the rich array of young stars around the major HII complexes,
leading to classification as a dwarf starburst or blue compact dwarf
\citep{ric01}.
However, the demonstration of \citet{hun86} that the classification of star-forming
dwarfs can be strongly distance-dependent leads us to re\"{e}xamine IC~10's
status as a Local Group enigma.

We have imaged a central region of IC~10, avoiding the actively starbursting
area,
with the Advanced Camera for Surveys (ACS) on HST, using the
ACS equivalents of V and I filters to produce CMDs of sufficient depth
to determine the SFH with timing precision of better than 10\% to ages
of $\approx$2000 Myr and $\sim$30\% over the entire lifetime of the 
galaxy \citep{col10}.  In order to better reveal possible
previous bursts of star formation and the older stellar populations in 
general we purposely avoided the regions previously imaged with HST
\citep{hun01}.  The resulting rich dataset contains a wealth of information
about IC~10, as is typical for HST images of nearby dwarfs 
\citep[e.g.][]{tol09,dal09}.  Of particular interest, we have identified
a window in the main body of IC~10 that is virtually free of internal 
reddening as identified by the narrow color range of red giants and
confirmed by the low HI column density at that location \citep{wil98}.

\begin{figure}[h]
\begin{center}
\includegraphics[scale=0.5, angle=0]{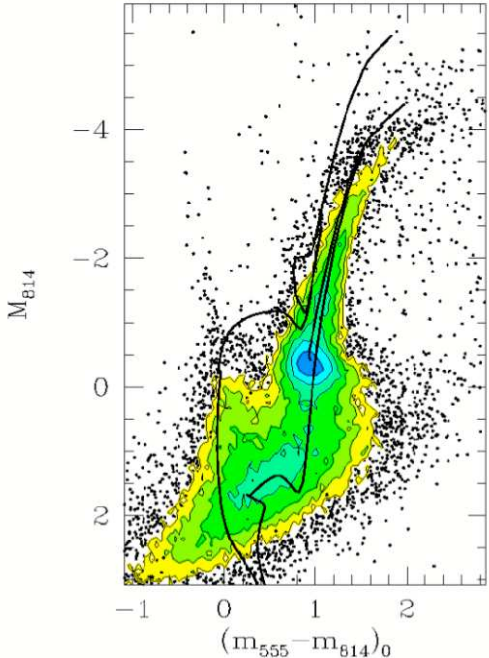}
\caption{A color-magnitude density diagram of a low-reddening window
in the northern disk of IC~10.  The negligible internal reddening permits
a reliable reconstruction of the SFH of the central regions of IC~10. 
The filters 814 and 555 are the ACS equivalents of I and V.
The CMD is remarkably similar to that of NGC~6822 \citet{wyd01}.
Isochrones are from the Padua group, computed for a metal abundance
Z = 0.004, with ages 400~Myr (upper track) and 2200~Myr (lower track).
5.6$\times$10$^4$ stars are measured in this 0.5~arcmin$^2$ window; the
contours are spaced by factors of 2 in density.
}\label{fig-ic10hess}
\end{center}
\end{figure}

The density-scaled CMD (Hess diagram) of 56,000 stars in this low-reddening
window is shown in Figure~\ref{fig-ic10hess}, where the magnitudes have
been corrected for a distance modulus of (m$-$M)$_0$ = 24.5 and a reddening
of E(B$-$V) = 0.81 mag.  Isochrones from \citet{mar08} have been overlaid
on the Hess diagram to show the locus of stars aged 400 and 2200~Myr with
metallicity Z = 0.004 ([M/H] $\approx$ 1/5th Solar, \citep{lee03}). 
This location in the galaxy is far from any sites of active star formation,
so the SFR for ages less than $\approx$1~Gyr is not expected to be 
representative of the galaxy as a whole.  For ages older than this, the
stellar velocity dispersion will have mixed stars throughout the body of 
the galaxy, so the SFH should be fairly typical of any random spot within
the central portion of IC~10.  The stellar population appears to be 
predominantly of intermediate age, with a strong contribution from stars
a few Gyr old.  

\begin{figure}[h]
\begin{center}
\includegraphics[scale=0.4, angle=0]{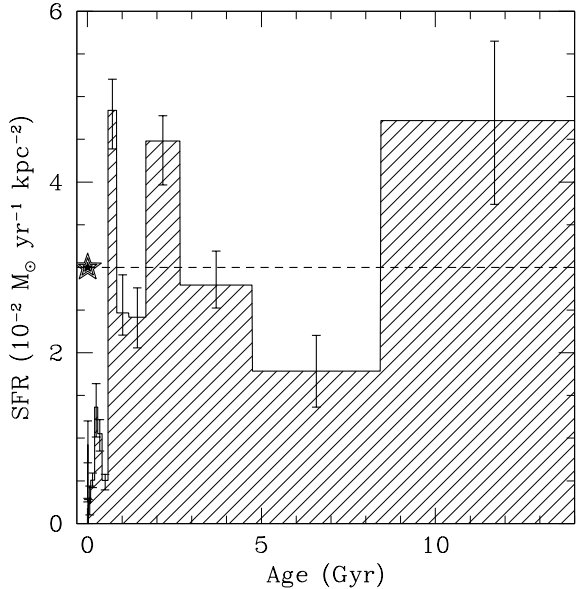}
\caption{The SFH of IC~10 determined in our low-reddening window.
For ages older than $\sim$1000~Myr orbital motions should have 
mixed IC~10's populations, so the SFH should be representative of the
galaxy as a whole for intermediate and old ages.  Note that by the
standards of its own historical average, IC10's current SFR
\citep[star and dashed line,][]{hun01} is not abnormal.
}\label{fig-ic10sfh}
\end{center}
\end{figure}

The SFH derived from this Hess diagram is shown in Figure~\ref{fig-ic10sfh}.
The lack of bright main-sequence stars in this specific field is reflected in 
the low SFR over the past $\approx$400~Myr.  However, this apparently quiescent
field far from the current center of starburst activity experienced its own
peaks in SFR at ages of approximately 700--800~Myr and 1.5--2.5~Gyr.  During
both time periods, the specific SFR at this location was higher than the 
current galaxy-wide specific SFR of 0.03~M$_{\odot}$~yr$^{-1}$~kpc$^{-2}$
\citep{hun01}, marked with a star at 10~Myr and extended for reference back
to 14~Gyr by the dashed line.  The Hess diagram shows that the age resolution
will be severely degraded for ages older than $\approx$3--4~Gyr, because the
main-sequence turnoff is too faint to be well-sampled by these data. 
However, it is apparent that the long-term average SFR of IC~10 may have approached
or exceeded its current value for much of its early history. 
This implies
that while IC~10's starburst is visually spectacular and prolific in production
of very massive stars, it is not likely to significantly increase the stellar
mass of IC~10 unless it increases in intensity or lasts for longer than a 
few hundred Myr.
One caveat is that
the current specific SFR from \citet{hun01} is averaged over the D$_{25}$ diameter
of the galaxy, an area $\approx$23 times the size of our low-reddening window, but the 
starburst activity is strongly concentrated into the high-surface-brightness
HII region complexes.  The peak current SFR, {\it measured over a comparable area to 
that shown here}, would thus be much higher than the D$_{25}$-averaged values.
Caution is warranted in making comparisons of this sort between studies of 
widely disparate areas.
However, the finite velocity dispersion of stellar populations coupled with the 
decreasing in time resolution of the CMD with age perform a sort of natural averaging
on the resolved stellar data, and this comparison should be valid for the older
ages considered, provided we have not had the misfortune to study a 
``special'' location in the galaxy where star formation has been unusually enhanced
or suppressed on Gyr timescales.

IC~10 has roughly the same stellar mass, dynamical mass, and average stellar
age as the prototypical dwarf irregular NGC~6822, the isolated dwarf 
irregular WLM, and the Milky Way companion Small Magellanic Cloud. In integrated
properties, there is little to distinguish between the four galaxies
\citep{lee03,dem04,orb08,col10}. The
similarity to NGC~6822 goes even farther, as both galaxies are
in the midst of
a minor merger or neutral gas accretion event \citep{hun97,wil98,deb06}.
The SFH for NGC~6822 determined by \citet{wyd01}
even bears similarity to the long-term average SFH for IC~10 presented in 
Figure~\ref{fig-ic10sfh}, characterized by roughly contant levels with 
evidence for a recent decline-- punctuated by enhancements or bursts that
may be connected to the presence of infalling gas.  The presence of dense
star clusters in NGC~6822 \citep[e.g.,][]{wyd01}, which are lacking in IC~10 
\citep{hun01,col10} suggests that if anything, IC~10 has had the 
{\it less} bursty SFH of the two.  It seems possible that if we viewed
IC~10 from a vantage point in M31 instead of through the plane of the Milky 
Way, we would think of IC~10 as a prototype dwarf irregular, with 
NGC~6822 as a distant analogue.

\section{Conclusions}

The burstiness of galaxies is a fundamental observable clue to understanding
the physics of star formation and the processes which drive galaxy evolution.
Starbursts, galaxies that are forming stars at such a high rate that the
background light of all previously formed stars pales to insignifcance 
\citep{san63}, are the extreme star formation environment in the Universe.
Starbursts were likely the dominant mode of star formation at high redshift 
\citep[][and references therein]{dre09}, but are not ``steady state'' 
phenomena in any sense \citep{rie75},
and their prevalence is low among luminous galaxies in the nearby Universe,
often tied to mergers and interactions \citep[e.g.,][]{ken98b}.

Late-type galaxies in the nearby Universe exhibit a wide range of 
specific star formation rates ranging from nearly inactive to 
extreme starburst conditions \citep{hun86,lee09}.  This is not surprising, as on 
theoretical grounds it is expected that galaxies with dynamical masses
M$_{tot} < \sim10^{9.5}$ M$_{\odot}$ become unstable to their own
stellar feedback from winds and supernovae \citep[e.g.,][and
references therein]{mac99,sti07}.  For small galaxies, this makes
the dichotomy between ``starbursting'' and ``quiescent'' states somewhat
artificial, because the natural state of dwarf galaxies is to show
a continuum of burstiness \citep{wei08} related to factors both internal (dynamical
mass, angular momentum profile, gas content) and external (tidal interactions,
ionizing background).  If a starburst is defined by a stellar birthrate
that exceeds 2--3 times the long-term average, then dwarf galaxies are 
much more susceptible to experiencing bursty star formation histories
than are giant galaxies.  While large galaxies most likely require
disruptive events and extreme conditions 
in order to experience a starburst, dwarf galaxies
should be able to meet the observational definition of a starburst without
qualitatively changing the mode of star formation, i.e., by producing
large numbers of massive clusters, concentrating the star formation to the
circumnuclear region of the galaxy, or experiencing major reorganizations of
gas content or morphology.

Within 10~Mpc, star formation `booms'' are rare, with only 
6\% of dwarf galaxies showing a
current SFR more than 2.5 times their long term average \citep{lee09}.  However,
``busts'' are equally rare, indicating that as long as neutral gas is present, 
some star formation occurs.  In general, the early conclusion of \citet{hun85}
suggesting that star formation in dwarf irregulars is ``down but not out'' has
been borne out and put on a firm statistical footing
by subsequent work.   Most recent work, typified by \citet{lee09}, suggests 
that between 20--30\% of star formation occurs during burst episodes.  
These episodes seem to last, very roughly, $\sim$10$^{8.5}$ yr 
\citep[e.g.,][]{mcq09}. Note that
these figures do not distinguish between systemic bursts, which are not
sustainable over the long term and may be signifcant in their production of
star clusters and metals or for their promotion of morphological transformation,
and stochastic bursts of the kind which are predicted by numerical models
\citep[e.g.,][]{sti07} and observed in nearby dwarfs \citep[e.g.,][]{wei08}.
It is interesting to note that while models predict a high degree of burstiness,
increasing with decreasing galaxy mass, CMD analysis indicates relatively
smooth SFHs for most Local Group galaxies.  Either some unmodelled factor is
acting to suppress the burstiness seen in the models, or the CMD analyses are
less sensitive than predicted to factors of 2 variation in SFR over time periods
of a few hundred Myr.

The isolated dwarf galaxy Leo~A appears to have waited several Gyr before forming 
the vast majority of its stars in an event that itself spanned several Gyr.
An upper limit of $\approx$10\% on the fraction of stars older than 8~Gyr was
found by \citet{col07}.  This makes it unique among galaxies that have been 
studied to the depth of the oldest main-sequence turnoff with the Hubble 
Space Telescope, but hints of similar behavior may be visible in other isolated
galaxies \citep[e.g.,][]{mcc06}.  If the delayed burst is a natural feature
of dwarf galaxies, then it has not been captured by the models that predict
steady ``breathing'' pulses of star formation.   One possibility is that 
cosmic reionization heated but did not evaporate the neutral gas from the
potential well of the galaxy and long cooling times produced the delay
\citep[e.g.,][]{bul00}.
On the other hand, if the 
late-blooming burst was triggered by a merger or accretion event, there is no
hope at this late time, several Gyr later, of identifying the trigger.

The galaxy IC~10 appears to have much in common with similar size dwarf irregulars
and is not currently forming stars far above its long-term average rate.  While
it has repeatedly been referred to as the nearest starburst galaxy and an anomaly
within the Local Group, the star formation history derived from HST imaging places
it squarely within a continuum of similar-mass late-type galaxies, bearing a strong
family resemblance to archetypal irregulars NGC~6822 and the Small Magellanic Cloud.
However the burst age is only a few times 10$^7$ yr, indicating that the 
burst may be at a very early stage if IC~10 has similar properties to other
dwarf starbursts, which appear to last for $\sim$10 times longer 
\citep[e.g.,][]{mcq09}.  IC~10 is similar in mass and metallicity to the 
SMC, but has failed to produce any massive star clusters during its lifetime, 
indirect evidence that the SMC has had the more tumultuous history and experienced
a more intense mode of star formation at past epochs.

\section*{Acknowledgments}
This article is a summary of a review talk presented at the
Southern Cross Astrophysics Conference, on the subject
of ``Galaxy Metabolism'', in June 2009.  The author gratefully
acknowledges the conference organisers for travel support, and the 
conference participants for lively and informative discussion
that influenced the form and content of this article.

\end{document}